\documentclass{appolb}
\usepackage{epsfig}
\usepackage{amssymb,amsmath}

\begin{document}
\title{High-$p_T$ Direct-Photon Results from PHENIX
\thanks{Presented at the PHOTON2005 International Conference on the 
Structure and Interactions of the Photon}%
}
\author{Klaus Reygers for the PHENIX Collaboration
\address{Institut f{\"u}r Kernphysik, University of M{\"u}nster \\
         Wilhelm-Klemm-Stra{\ss}e 9, 48149 M{\"u}nster, Germany}
}
\maketitle

\begin{abstract}
  Direct-photon measurements in p+p and Au+Au collisions at
  $\sqrt{s_\mathrm{NN}} = 200$\,GeV from the PHENIX experiment are
  presented. The p+p results are found to be in good agreement with
  next-to-leading-order (NLO) perturbative QCD calculations.
  Direct-photon yields in Au+Au collisions scale with the number of
  inelastic nucleon-nucleon collisions and don't exhibit the strong
  suppression observed for charged hadrons and neutral pions. This
  observation is consistent with models which attribute the
  suppression of high-$p_\mathrm{T}$ hadrons to energy loss of quarks
  and gluons in the hot and dense medium produced in Au+Au collisions
  at RHIC.
\end{abstract}
  
\section{Direct Photons in p+p Collisions}
\begin{figure}[t]
\centerline{\epsfig{file=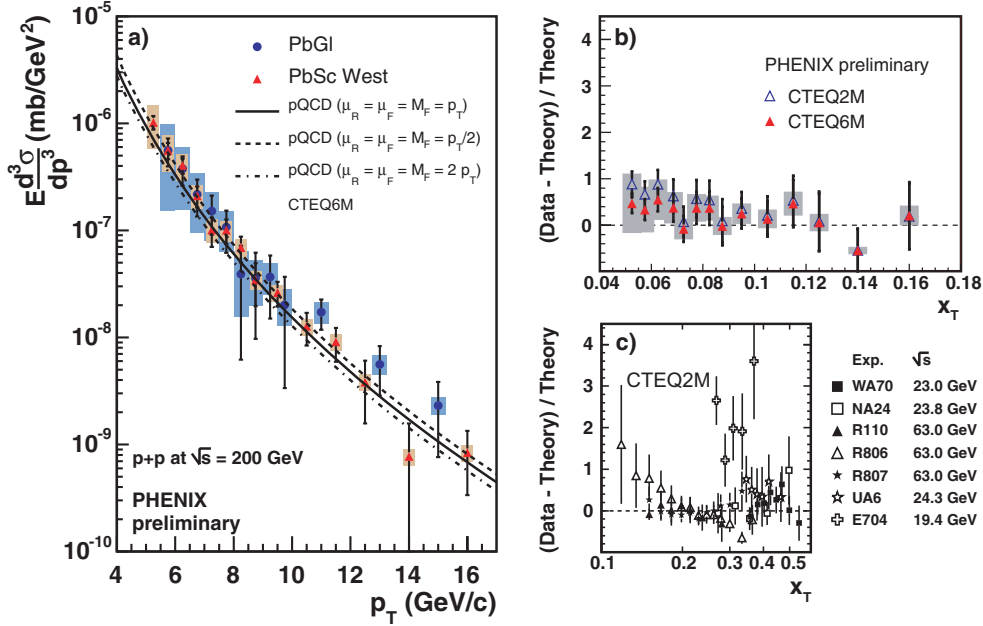,width=13cm}}
\caption{a) Invariant direct-photon cross-section in p+p collisions 
  at $\sqrt{s} = 200$~GeV determined with the lead scintillator (PbSc)
  and the leadglass (PbGl) calorimeter of the PHENIX experiment. b)
  Comparison of the PbSc direct-photon spectrum with two pQCD
  calculations which employ the CTEQ2M and the CTEQ6M parton
  distribution function, respectively. c) Comparison of direct-photon
  world data for p+p collisions at various energies with pQCD
  calculations which employ the CTEQ2M parton distribution function
  \cite{Vogelsang:1997cq}.}
\label{fig:cross_sec_pp}
\end{figure}

The virtue of direct photons is that they emerge directly from the
interaction of point-like partons in processes like quark-gluon
Compton scattering ($\mathrm{q}+\mathrm{g} \rightarrow
\mathrm{q}+\gamma$) and quark-anti-quark annihilation
($\mathrm{q}+\bar{\mathrm{q}} \rightarrow \mathrm{g}+\gamma$).
Measurements of direct photons in p+p collisions therefore allow to
test perturbative QCD (pQCD) without the complication of
phenomenological parton-to-hadron fragmentation functions which are
needed in the description of high-$p_\mathrm{T}$ hadron production.
Moreover, direct-photon measurements in p+p collisions provide
information about the gluon distribution of the proton due to the
large contribution from quark-gluon Compton scattering. This is
especially interesting for fractional gluon momenta
$x_\mathrm{Bjorken} \gtrsim 0.1$ because this range is not well
constrained by other processes like Deep-Inelastic Scattering and
Drell-Yan. However, direct-photon data from $\mathrm{p}+\mathrm{p}$
and $\mathrm{p}+\bar{\mathrm{p}}$ collisions are not generally used in
global QCD fits for the determination of parton distribution
functions. One reason for this are discrepancies between data and
pQCD, especially at low energies (see e.g.
Figure~\ref{fig:cross_sec_pp}c).  The Relativistic Heavy Ion Collider
(RHIC) delivers spin polarized proton beams. A further motivation for
direct-photon measurements in PHENIX is to gain insight into the
contribution of gluons to the spin of the proton by comparing
direct-photon yields in p+p collisions with equal and opposite proton
helicities. In this paper, however, the direct-photon cross-section
for unpolarized p+p collisions is presented.  The experimental
challenge in direct-photon measurements is the subtraction of the
significant photon background from hadron decays like $\pi^0
\rightarrow \gamma+\gamma$ and $\eta \rightarrow \gamma+\gamma$.

The comparison of world data on direct-photon production in
$\mathrm{p}+\mathrm{p}$ and $\mathrm{p}+\bar{\mathrm{p}}$ collisions
with pQCD calculations raises the question whether there is a
systematic pattern of deviation. It has been argued that measured
direct-photon transverse momentum ($p_\mathrm{T}$) spectra generally
tend to be steeper than pQCD predictions \cite{Huston:1995vb}. In this
case the agreement between data and pQCD can be improved by
introducing a transverse momentum ($k_\mathrm{T}$) of the partons
prior to the hard scattering, in addition to the effects expected in
the next-to-leading-order description. Recently published results on
direct-photon production in p+p and p+Be collisions from the
fixed-target experiment E706 provide further support for the idea that
$k_\mathrm{T}$ enhancement needs to be taken into account to describe
the data \cite{Apanasevich:2004dr}.

Direct-photon spectra in $\sqrt{s_\mathrm{NN}} = 200$\,GeV p+p
collisions were measured at RHIC with the highly segmented
electromagnetic calorimeter (EmCal) of the PHENIX experiment
\cite{Okada:2005in}.  This detector consists of two subsystems: a lead
scintillator sampling calorimeter (PbSc) and a lead glass Cherenkov
calorimeter (PbGl). The EmCal covers the pseudorapidity range $|\eta|
< 0.35$. The direct-photon cross-section for unpolarized p+p
collisions determined with data from RHIC Run-3 is shown in
Figure~\ref{fig:cross_sec_pp}.  The integrated luminosity for this run
was 266\,nb$^{-1}$. The independent PbSc and PbGl measurements are
consistent and agree with NLO pQCD predictions.  There is, in
particular, no indication that for the range $p_\mathrm{T} \gtrsim
5$\,GeV/$c$ covered by the current measurement an additional
transverse momentum $k_\mathrm{T}$ is needed in the theoretical
description.
\begin{figure}[t]
\centerline{\epsfig{file=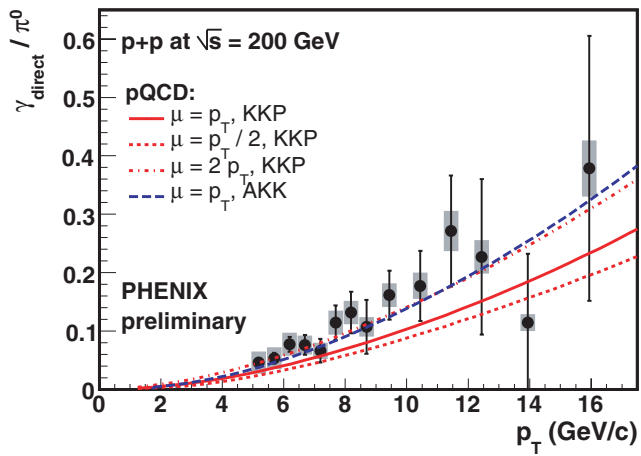,height=6cm}}
\caption{Comparison of the ratio of the direct-photon and neutral-pion invariant 
  cross-sections for p+p collisions at $\sqrt{s} = 200$~GeV with pQCD
  predictions.  The calculation for the KKP parton-to-pion
  fragmentation functions \cite{Kniehl:2000fe} is shown for three
  choices $\mu = \mu_\mathrm{R} = \mu_\mathrm{f} = M_\mathrm{F}$ of
  the renormalization scale $\mu_\mathrm{R}$, the initial-state
  factorization scale $\mu_\mathrm{f}$, and the final-state
  factorization scale $M_\mathrm{f}$. Moreover, a calculation is shown
  which employs the AKK fragmentation functions \cite{Albino:2005me}.}
\label{fig:gam_pi0_ratio}
\end{figure}

Experimental and theoretical uncertainties are potentially reduced in
the ratio of the direct-photon and neutral-pion cross-sections.
Figure~\ref{fig:gam_pi0_ratio} shows that the measured
$\gamma_\mathrm{direct}/\pi^0$ ratio agrees with NLO pQCD predictions.
The AKK fragmentation functions \cite{Albino:2005me} provide a better
description of the PHENIX $\pi^0$ spectrum \cite{Adler:2003pb} than
the KKP fragmentation functions and this in turn leads to a better
agreement between the measured and calculated
$\gamma_\mathrm{direct}/\pi^0$ ratio.

\section{Direct Photons in Au+Au Collisions}
Direct-photon measurements in nucleus-nucleus (A+A) collisions
essentially serve two purposes. At sufficiently high transverse
momenta direct photons emerge from initial hard parton-parton
scattering, analogous to the production mechanism in p+p collisions.
These hard parton-parton scatterings happen prior to the possible
formation of a thermalized quark-gluon plasma (QGP). However, due to
their electromagnetic nature photons essentially don't interact with
the hot and dense plasma.  They can thus be used as a measure of the
number of initial hard parton-parton scatterings. The second incentive
to measure direct photons in A+A collisions is the expected production
of thermal direct photons in the QGP. Thermal direct photon are
predominantly produced in the early hot phase of the evolution of the
plasma and thus provide a means to determine the initial temperature
of the QGP. In a window $1 \lesssim p_\mathrm{T} \lesssim 3$\,GeV/$c$
thermal direct photons from the QGP are expected to be the dominant
direct-photon source in Au+Au collisions at $\sqrt{s_\mathrm{NN}} =
200$\,GeV.

One of the most important observations at RHIC is the suppression 
high-$p_\mathrm{T}$ hadrons relative to a scaled p+p reference.
The suppression is quantified with the nuclear modification factor 
\begin{equation}
R_\mathrm{AA}(p_\mathrm{T}) = 
  \frac{\mathrm{d}N/\mathrm{d}p_\mathrm{T}|_\mathrm{A+A}}
       {\langle T_\mathrm{AA}\rangle 
        \times \mathrm{d}\sigma/\mathrm{d}p_\mathrm{T}|_\mathrm{p+p}}.
\end{equation}
The factor $\langle T_\mathrm{AA} \rangle = \langle
N_\mathrm{coll}\rangle / \sigma_\mathrm{inel}^\mathrm{p+p}$ reflects
the increase of the parton luminosity per A+A collision relative to
p+p collisions in the absence of nuclear shadowing. Thus, in the
absence of initial-state or final-state nuclear effects
$R_\mathrm{AA}$ will be unity for sufficiently large transverse
momenta for which particle production is dominated by hard scattering.
The average number $\langle N_\mathrm{coll} \rangle$ of inelastic
nucleon-nucleon collisions for a given centrality class is determined
with a Glauber model calculation.

\begin{figure}[t]
\centerline{\epsfig{file=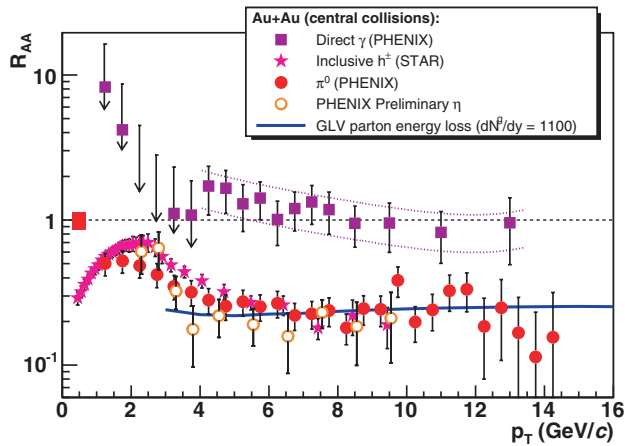,height=6cm}}
\caption{Nuclear modification factor $R_\mathrm{AA}$ for direct photons 
  and hadrons in central (0-10\% of
  $\sigma_\mathrm{inel}^\mathrm{Au+Au}$) Au+Au collisions at
  $\sqrt{s_\mathrm{NN}} = 200$~GeV.}
\label{fig:raa}
\end{figure}

As can seen in Figure~\ref{fig:raa} neutral pions, $\eta$-mesons, and
charged hadrons are suppressed by a factor $\sim 5$ in central Au+Au
collisions at $\sqrt{s_\mathrm{NN}} = 200$\,GeV. This can be explained
by energy loss of fast quarks and gluons in the QGP.  In such models
the initial rate of hard parton-parton scatterings scale as $\langle
T_\mathrm{AA} \rangle$ and the suppression is only due to the
final-state interaction with the medium. The measurement of direct
photons at high $p_\mathrm{T}$ provides a unique possibility to test
these jet quenching models \cite{Adler:2005ig}. Figure~\ref{fig:raa}
shows that the $R_\mathrm{AA}$ for direct photons in central Au+Au
collisions is consistent with unity: unlike hadrons, direct photons
are not suppressed. One can conclude that initial hard scattering
processes in Au+Au collisions indeed occur at the rate expected from
$\langle T_\mathrm{AA} \rangle$ scaling. The suppression of hadrons at
high $p_\mathrm{T}$ is therefore a final-state effect caused by the
hot and dense medium produced in these collisions. This supports
models which attribute the hadron suppression to energy loss of
partons in the medium and is consistent with the formation of a QGP in
Au+Au collisions at RHIC.

{\it Acknowledgment}. We would like to thank Werner Vogelsang for
providing the QCD calculations used in this paper.

\end{document}